\documentclass[twocolumn]{aastex63}
\usepackage{hyperref,amssymb,amsmath,graphicx}
\shorttitle{Growth of stellar mass black holes via Bondi accretion}
\shortauthors{Rice and Zhang}
\begin{document}
\title{Growth of stellar mass black holes in dense molecular clouds and GW190521}

\author[0000-0003-3887-091X]{Jared R. Rice}
\affiliation{Department of Physics, Texas State University, San Marcos, TX 78666, USA}
\email{jrice@txstate.edu}

\author[0000-0002-9725-2524]{Bing Zhang}
\affiliation{Department of Physics and Astronomy, University of Nevada Las Vegas, Las Vegas, NV 89154, USA}
\email{zhang@physics.unlv.edu}

\begin{abstract}
A stellar mass black hole can grow its mass noticeably through Bondi accretion, if it is embedded in an extremely dense and massive molecular cloud with slow motion with respect to the ambient medium for an extended period of time. This provides a novel, yet challenging channel for the formation of massive stellar-mass black holes. We discuss how this channel may account for the massive binary black hole merger system GW190521 as observed by LIGO/Virgo gravitational wave detectors as well as the claimed massive black hole candidate LB-1.
\end{abstract}

\section{Introduction} \label{sec:intro}
Recently, the LIGO/Virgo Scientific Collaboration reported the discovery \cite{Abbott20a} of a binary black hole (BH) merger gravitational wave event, GW190521. The masses of the two pre-merger BHs are $85^{+21}_{-14} M_\odot$ and $66^{+17}_{-18} M_\odot$, respectively, and the mass of the final BH is $142^{+28}_{-16} M_\odot$. The probability that at least one of the merging black holes is in the mass range of the so-called pulsational pair-instability supernova gap ($65-120 M_\odot$) is 99.0\%, suggesting that such systems are not easy to form based on standard stellar evolution models. \cite{Abbott20b} discussed several possible astrophysical channels for forming such massive binary BH systems, such as hierarchical mergers (in globular clusters), stellar mergers (direct collapse of merged stars with an oversized hydrogen envelope), mergers in AGN disks, as well as primordial BHs. They also disfavored the possibility of strong gravitational lensing for GW190521.  \cite{DeLuca2020} proposed that the BH masses observed in GW190521 may be explained by single or binary primordial black hole systems that have undergone a period of early (pre-reionization) accretion. \cite{Mohammadtaher2020} suggested that the BH is a remnant of a Population III star, accreting ambient high-density gas inside of a high-redshift minihalo. \cite{Natarajan2020} suggested that the BH accretes gas inside of a nuclear star cluster. Previous works on single and binary black hole accretion, such as \citep{Roupas2019a,Roupas2019b}, have suggested that massive black holes with masses in the pair-instability gap or lower may form via accretion  inside star clusters. More exotic explanations for the massive black holes observed in GW190521 may be found in \cite{Sakstein2020}, where several possible explanations beyond the Standard Model are discussed. An in-depth account of black hole accretion inside nuclear clusters is discussed in \cite{Kroupa2020}.

Another massive stellar-mass BH candidate is in a binary system called LB-1, which was initially reported to have a mass $M = 68_{-13}^{+11}\,M_\odot$ \citep{Liu19}. If this mass is true\footnote{This suggested mass was questioned by some groups \citep[e.g.][]{AbdulMasih20,Shenar2020,Eldridge20}. Recent updated analyses by the original group \citep{Liu20,Liu20b} derived a lower BH mass for LB-1.}, then its existence also challenges the standard stellar evolution theories.

Here we propose an alternative astrophysical channel to form massive stellar-mass BHs. We suggest that if molecular clouds (MCs) are formed in the regions where stellar-mass BHs (from the deaths of the previous generation of massive stars) are located, the BHs may undergo significant growth through accretion from the ambient gas under certain conditions. We derive these conditions and discuss the possible formation of the GW190521 progenitor system and LB-1 within such a framework. 

\section{Black holes in molecular clouds}

Isolated BHs in MCs have been discussed in the past, mostly within the context of radiation signatures that reveal their existence \citep{McDowell85,Campana93}. \cite{Campana93} estimated that each generation of stars produce $\sim 3.5 \times 10^{-4} \, M_\odot^{-1}$ BHs. MCs have a typical diameter $\sim 40$ pc and mass $10^5-10^6 M_\odot$. Considering a filling factor of MCs in the Milky Way, they estimated there are $\sim 2$ detectable accreting BHs per MC. In these previous studies, the growth of BHs was not investigated.

Following our previous study on the growth of primordial BHs \cite{RiceZhang17}, we study the possible growth of BHs embedded in MCs within the lifetime of MCs. \cite{Jeffreson2018,Jeffreson2020} proposed an analytical model of MC lifetimes that incorporates galactic dynamics. In their model (see \cite{Jeffreson2018} Eq. 24), the lifetime of a molecular cloud is the sum
\begin{align}
    \tau &= \bigl|\tau_\kappa^{-1} + \tau_{\Omega_\text{P}}^{-1} + \tau_\text{ff,g}^{-1} + \tau_\text{cc}^{-1} - \tau_\beta^{-1} \bigr|^{-1},\label{eq:lifetime}
\end{align}
\noindent where $\tau_\kappa$ is the epicyclic perturbation timescale, $\tau_{\Omega_\text{P}}$ is the spiral arm crossing timescale, $\tau_\text{ff,g}$ is the ISM freefall timescale, $\tau_\text{cc}$ is the cloud-cloud collision timescale, and $\tau_\beta$ is the galactic shear timescale. In this analytical model, the predicted MC lifetimes vary with the galactocentric radius. For the Milky Way, their prediction is that the MC lifetimes should vary between $\sim 20$ Myr and $\sim60$ Myr. Beyond the Solar neighborhood at distances greater than $\sim 8.0$ kpc, the predicted MC lifetime increases to its maximum value. In our model the BHs exist prior to the formation of the MCs, so one may use $20-60$ Myr as the duration for BH mass growth.

\section{Black hole accretion} \label{sec:accretion}

We consider quasi-spherical accretion of BHs in MCs for two reasons. This allows us to estimate the accretion rate using the simplest Bondi-Hoyle-Lyttleton (hereafter BHL; see \cite{Bondi52}) rate. Also a low angular momentum cloud can avoid an extended accretion disk so that the inflow can maintain a low sound speed. Numerical simulations \citep{waters20} show that inhomogeneous distributions of rotational velocity and density along the outer radius marking the sphere of influence of a BH would modify Bondi-like accretion to form outflows along with inflows, but with the accretion rate remaining close to the Bondi rate value.  Simulations of choked black hole accretion with the accretion rate remaining near the Bondi rate value were shown in  \cite{Aguayo-Ortiz2019,Aguayo-Ortiz2020,Tejeda2020}.

One factor that will slow down BH growth is the feedback effect. For example, \cite{Krolik2004} proposed a model to explain ultraluminous X-ray sources (ULXs) as accreting intermediate mass BHs. In his model, the BHs accrete at the standard BHL rate, but since radiative feedback significantly reduces the local number density and increases the ambient ISM temperature, the actual accretion rate is reduced by a few orders of magnitude. \cite{Park2013} performed 2-D axisymmetric hydrodynamical simulations to arrive at an analytical formulation of BHL accretion, regulated by radiative feedback, in the case of a moving BH. They found that for a BH supersonically traversing a high density medium, the accretion rate onto the BH is steady. Interestingly, if the BH's relative velocity is about a few times the sound speed, the accretion rate maintains or even rises above the Bondi rate \citep{Bondi52} even if radiative feedback is considered. This is because the dense ionization front and bow shock decrease the downstream relative gas velocity. Additionally, \cite{Park2013} discovered that the accretion rate in this parameter regime is independent of the ambient ISM temperature so that the growth rate of moving BHs is actually higher than that for stationary BHs. In our problem, the MCs have low sound speeds (see Eq.(\ref{eq:cs}) below), so that the supersonic condition in the \cite{Park2013} simulations is readily satisfied.

In the following, we adopt the following MC fiducial parameters: electron number density $n = (10^5\,  {\rm cm^{-3}})\, n_5$, mass $M_{mc} = (10^6\, M_\odot)\, M_{mc,6}$, so that the radius of the cloud is $R_{mc} = [(3/4\pi) M_{mc}/nm_p]^{1/3} = (4.6\,  {\rm pc})\, M_{mc,6}^{1/3}\, n_5^{-1/3}$, where $m_p$ is the proton mass. In general, the accretion rate of an isolated BH can be expressed as
\begin{align}
\frac{dM}{dt} &= \min\bigl(\dot{M}_{\rm BHL},\dot{M}_{\rm E}\bigr),
\end{align}
which is bound by the Eddington accretion rate
\begin{align}
\nonumber\dot{M}_{\rm E} &= \frac{4\pi Gm_p}{\varepsilon\sigma_Tc}M\\
&\simeq (2.2\times 10^{-7}\,M_\odot \, {\rm yr}^{-1}) \
\varepsilon_{-1}^{-1}\biggl(\frac{M}{10\,M_\odot}\biggr),\label{eq:Eddi}
\end{align}
where $G$ is Newton's gravitation constant, $\sigma_T$ is the Thomson scattering cross section for electrons, $M$ is the mass of the BH, $c$ is the speed of light, and $\varepsilon$ denotes the efficiency of converting the accretion power to luminosity, which has a typical value of $\sim 0.1$. The BHL accretion rate is
\begin{eqnarray}
\dot{M}_{\rm BHL} &=& \frac{4\pi\lambda_c G^2}{V^3}m_pn M^2
\nonumber \\
&\simeq & (1.5\times 10^{-8}\,M_\odot \, {\rm yr}^{-1}) \ \biggl(\frac{V}{10 \, {\rm km \, s^{-1}} }\biggr)^{-3}
\nonumber \\
& \times & n_5\biggl(\frac{M}{10\,M_\odot}\biggr)^2,\label{eq:Bonfin}
\end{eqnarray}
\noindent where 
\begin{equation}
V\equiv (c_s^2+v^2)^{1/2},
\end{equation}
and
\begin{equation}
c_s = \biggl(\frac{\hat\gamma k_BT}{m_p}\biggr)^{1/2} \simeq (1.2 \, {\rm km \, s^{-1}})\,T_2^{1/2}
\label{eq:cs}
\end{equation}
is the sound speed in the MC far from the hole, $v$ is the relative velocity between the BH and the ambient gas, $\lambda_c =1/4$ (if $\hat{\gamma}=5/3$) is the non-dimensional accretion eigenvalue given in \cite{Bondi52}, $\hat\gamma=5/3$ is the adiabatic index, $k_B$ is the Boltzmann constant, and $T=100\,\text{K} \ T_2$ is the medium temperature far from the hole.

The temperature range of MCs is ${T\sim(10-10^2)\,\text{K}}$, which corresponds to a sound speed range of $c_s \sim (0.4-1.2) \, { \rm km \, s^{-1}}$. The proper motion velocity $v$ of the BH is likely greater than $c_s$, so the BHL accretion rate is mainly defined by $V \sim v \gg c_s$. Hereafter our nominal parameter is $v \sim 10 \, {\rm km \, s^{-1}}$. From the scaling in Eq.(\ref{eq:Bonfin}), we immediately infer that a relatively massive BH (say, $M > 10 M_\odot$) embedded in a MC could grow significantly within the lifetime of the MC if $V$ is small enough and $n$ is large enough. Comparing Eqs.(\ref{eq:Eddi}) and (\ref{eq:Bonfin}), we also conclude that the Eddington limit does not kick in as long as the BH mass satisfies
\begin{equation}
    M < 150\, M_\odot\, \varepsilon_{-1}^{-1} n_5^{-1} \biggl(\frac{V}{10 \, {\rm km \, s^{-1}} }\biggr)^{3}.
\end{equation}

A supersonic BH will remain inside a MC as long as the escape velocity of the MC is higher than the BH velocity. The escape velocity of our nominal MC is
\begin{align}
    v_\text{esc} &= \sqrt{\frac{2GM_{mc}}{R_{mc}}}\nonumber\\
    &= 43\,{\rm km \, s^{-1}}
    M_{mc,6}^{1/2}
    \biggl(\frac{R_{mc}}{4.6 \,{\rm pc}}\biggr)^{-1/2},
\end{align}
Our nominal BH is traveling at $v=10\,{\rm km \, s^{-1}}$, so  $v<v_\text{esc}$ and the BH remains within the MC for the entire lifetime of the MC as it grows its mass. Notice that we envisage a scenario where the MC is formed at the location of BHs formed from the previous generation stars. If the BH is captured by the MC, its initial speed should exceed the escape velocity. A significant mass growth is possible only if the BH can slow down due to dynamical friction and remain within the MC to gain mass.

\begin{figure*}[t]
\begin{center}
\includegraphics[width=\columnwidth]{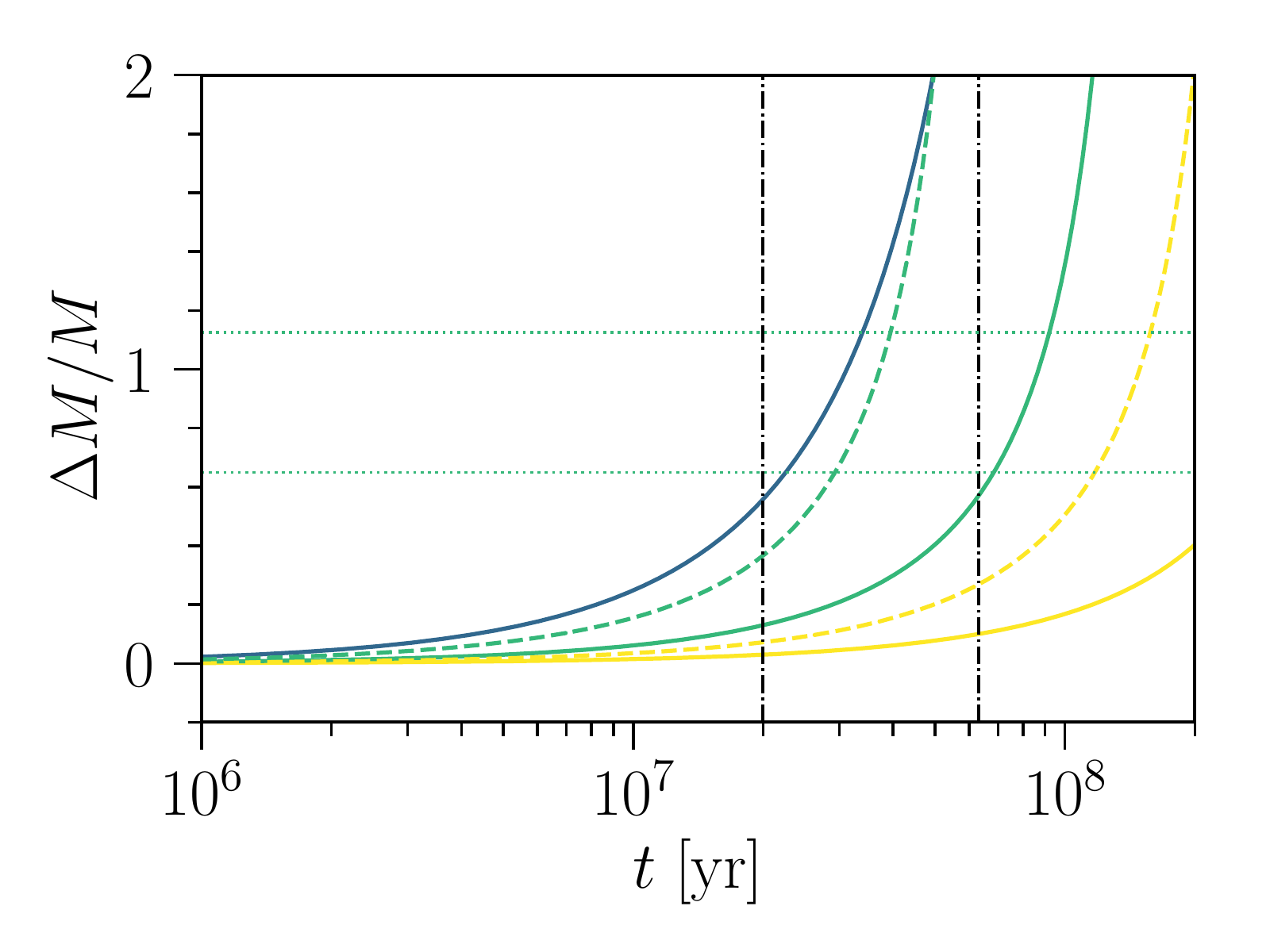}
\includegraphics[width=\columnwidth]{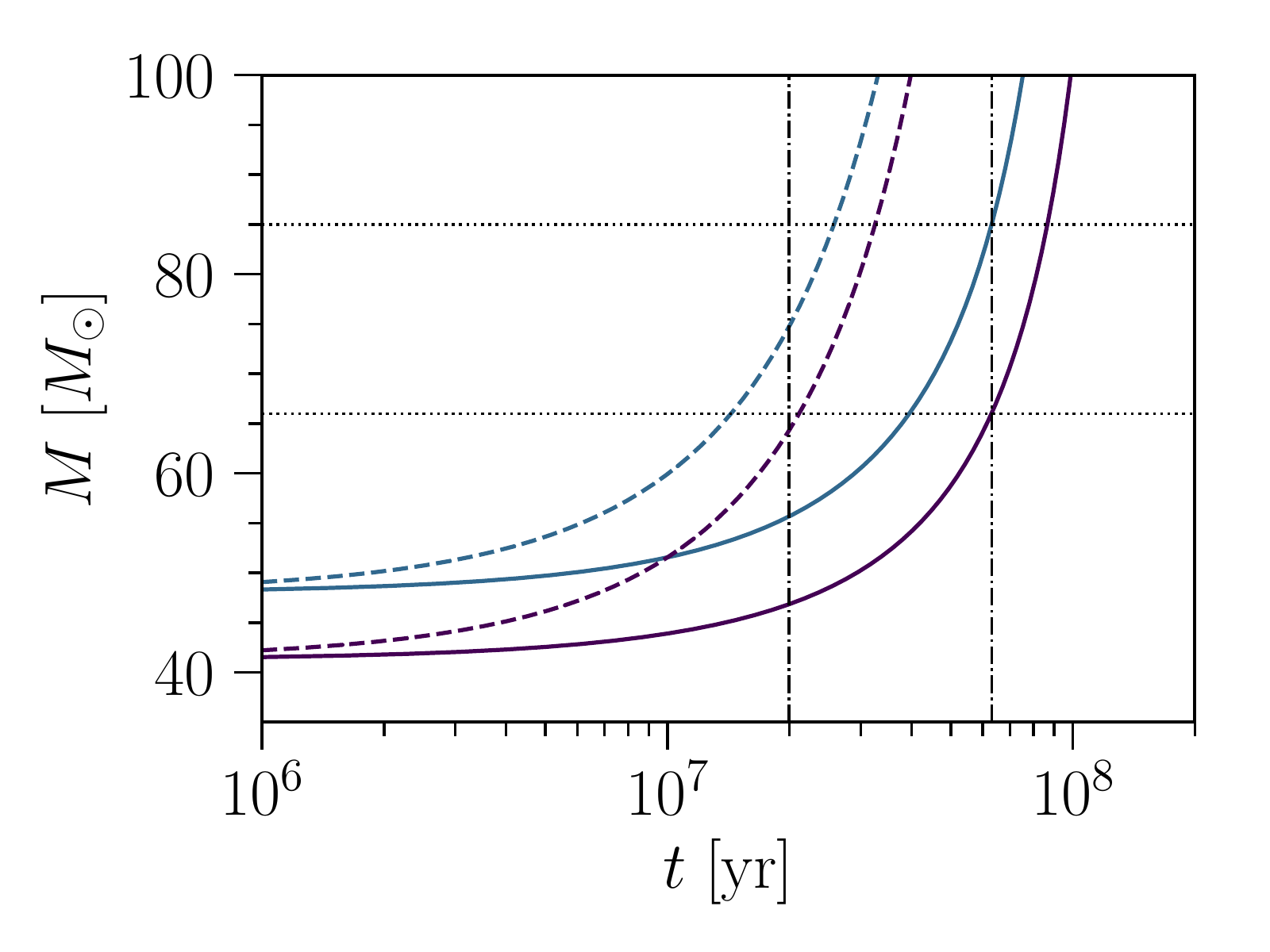}
\caption{\label{fig:deltaM}{\bf Left}: Fractional mass increase of Bondi-accreting BHs embedded in a molecular cloud, as a function of time. The two dot-dashed vertical lines represent upper and lower bound estimates from \cite{Jeffreson2018} on the lifetime of MCs in the Milky Way, from $t=19.9\text{ Myr}$ to $t=63.1\text{ Myr}$. In each case, the BH is embedded in a molecular cloud with the following physical parameters: $n=10^5\text{ cm}^{-3}$, $T=10^2\text{ K}$, and $v=10.0\text{ km$\,$s$^{-1}$}$ (solid) or $v=7.5\text{ km$\,$s$^{-1}$}$ (dashed). {\bf Blue}: a BH accreting at the Eddington rate given in Eq. \ref{eq:Eddi}. {\bf Green}: a BH of mass $M=40\,M_\odot$ accreting at the BHL rate given in Eq. \ref{eq:Bonfin}. {\bf Yellow}: a BH of mass $M=10\,M_\odot$ accreting at the BHL rate given in Eq. \ref{eq:Bonfin}. The two green dotted horizontal lines represent the fractional mass gain of $\Delta M/M = 45/40 = 1.125$ (top) and $\Delta M/M = 26/40 = 0.65$ (bottom) of a $40\,M_\odot$ BH accreting to the estimated BH masses in the GW190521 system (see \cite{Abbott20a}) of $85\,M_\odot$ and $66\,M_\odot$, respectively. {\bf Right}: Mass (in solar masses) of Bondi-accreting BHs embedded in a molecular cloud as a function of time. The two dot-dashed vertical lines are the same as in the left panel. The two dotted horizontal lines represent the estimated GW190521 masses of $85\,M_\odot$ and $66\,M_\odot$ from \cite{Abbott20a}. The solid lines represent the evolution track of two BHs with initial critical masses that allow the masses to grow to the GW190521 masses within the upper-bound timescale $t=63.1\text{ Myr}$. The initial masses of such BHs are $41.3\,M_\odot$ ({\bf purple}) and $48.0\,M_\odot$ ({\bf blue}), respectively. The dashed lines represent BHs with those initial masses that are accreting at the Eddington rate.}
\end{center}
\end{figure*}

Figure \ref{fig:deltaM} left panel shows the growth of BHs with initial masses $m_i = 10\,M_\odot,\, 40\,M_\odot$ embedded in a MC with $n_5=1$. The two black dashed vertical lines bracket the range of the lifetime of a typical MC from \cite{Jeffreson2018}. One can see that a relatively massive BH can undergo significant growth if $v$ is small enough. For our nominal parameters ($n=10^5\,\text{cm}^{-3}$, $T=10^2\,\text{K}$, and $v=10.0\,\text{km$\,$s$^{-1}$}$), a black hole with initial mass $m_i=10\,M_\odot,\, 40\,M_\odot$ would gain a fractional mass $\Delta M/M \sim 10\%,\,57\%$, respectively. A larger $M_i$ or a smaller $v$ would enhance the mass growth of the BH. This provides a novel channel for producing massive stellar-mass BHs.

In our calculation, we have assumed that the MC is not destroyed due to radiative feedback throughout the epoch of BH growth. Even though radiative feedback was found not to be important locally in the simulations of \cite{Park2013}, one still needs to justify this over a much longer timescale. With our fiducial parameters, the total binding energy of the MC is 
\begin{equation}
 E_b = \frac{G M_{mc}^2}{R_{mc}} \simeq (1.9\times 10^{52} \, {\rm erg}) \, M_{mc,6}^2 \left( \frac{R_{mc}}{4.6 \ {\rm pc}}\right)^{-1}. 
\end{equation}
The dissociation energy per molecule of H$_2$ is $4.5\,{\rm eV}$ and the ionization energy per atom of H is $13.6\,{\rm eV}$, so one can estimate the total energy needed to break up molecules
\begin{equation}
    E_{\rm mol} = \frac{M}{2 m_p} \, 4.5 \, {\rm eV} \simeq 4.3 \times 10^{51} \, {\rm erg} \, M_{mc,6},
\end{equation}
and the energy needed to break up atoms
\begin{equation}
    E_{\rm atm} = \frac{M}{m_p} \, 13.6 \, {\rm eV} \simeq 2.6 \times 10^{52} \, {\rm erg} \, M_{mc,6}.
\end{equation}
As a result, the total energy to destroy a $10^6 M_\odot$ MC is
\begin{equation}
    E_{\rm tot} = E_b + E_{\rm mol} + E_{\rm atm} \sim 4.9 \times 10^{52} \, {\rm erg}.
\end{equation}
With the Eddington luminosity $L_{\rm Edd} = 1.3\times 10^{39} \, {\rm erg \, s^{-1}} (M/10\, M_\odot)$ of the accreting black hole, the time it takes to destroy the MC is\footnote{This estimate is likely an upper limit, since depositing heat to the MC would alter the structure of the MC, which may decrease the accretion rate. In any case, $\dot M_{\rm BHL}$ in our calculation (Eq.(\ref{eq:Bonfin})) is dominantly defined by the proper motion velocity $V \sim 10 \, {\rm km \, s^{-1}}$, which is one order of magnitude greater than the sound speed for fiducial parameters. The results would not change significantly unless the temperature of cloud reaches the order of $10^4$ K.}
\begin{equation}
    t_{\rm destroy} = E_{\rm tot}/L_{\rm Edd} = 1.2 \times 10^6 \, {\rm yr}.
\end{equation}
This is shorter than the BH mass growth time (20-60 Myr) in our calculation, but is much longer than the timescale $t_{\rm esc}$ for the radiation to escape the MC. The latter can be calculated as
\begin{equation}
    t_{\rm esc} \simeq \frac{R_{mc}}{c}\, \tau = 2.1 \times 10^4 \, {\rm yr} \,
    n_5 \sigma_{-21} \left(\frac{R_{mc}}{4.6 \, {\rm pc}}\right)^2,
\end{equation}
where 
\begin{equation}
    \tau = n \sigma R_{mc} \simeq 1.4\times 10^3\,n_5 \sigma_{-21}\, \left(\frac{R_{mc}}{4.6 \, {\rm pc}}\right).
\end{equation}
is the optical depth of X-ray photons from the disk, with the absorption cross section $\sigma \sim 10^{-21} \, {\rm cm^2} \,\sigma_{-21}$ \citep{wilms00} normalized to the cross section at $\sim 0.5$ keV, roughly the typical photon energy of a Shakura-Sunyaev accretion disk around a $\sim 10\,M_\odot$ BH \citep{shakura73}. If the photon energy is lower (e.g. by getting reprocessed by absorption and re-emission), the cross section would be higher, which would increase the photon trapping time. The MC may be then destroyed to halt BH growth. As a result, radiative feedback would not be important only under extreme conditions. It can be indeed important for most MCs with less extreme parameters (e.g. lower $n$ and lower $M_{mc}$), so BHs can grow significantly in MCs only in a small parameter space. This is consistent with the rarity of massive BHs in the MW. Further detailed numerical simulations are needed to verify our suggestion for the fiducial parameters adopted in this paper.

\section{Case studies} \label{sec:cbc}
\subsection{GW190521}

Both pre-merger BHs in GW190521 are massive ($85^{+21}_{-14}\,M_\odot$ and $66^{+17}_{-18}\,M_\odot$). The hierarchical BH merger scenario  \citep[e.g.][]{fragione20} would require that each of the two components has undergone one merger in stellar clusters prior to the detected merger. Within that scenario, one would expect some other merging systems should have one massive component (after one merger) and one less massive component (without merger yet). In our scenario, the two BHs likely grow in the same MC and gain mass simultaneously. It is natural to obtain two heavy BHs in one merging system. Figure \ref{fig:deltaM} right panel shows that for our nominal parameters, two BHs with initial masses $41.3 M_\odot$ and $48 M_\odot$ (both consistent with formation from massive, especially first generation, stars; \citealt{heger03}) could grow to masses consistent with GW190521 within the lifetime of a dense MC. 

The requirement for significant mass growth is relatively slow motion between the accreting BH and the MC. For a binary system with a total mass $m \equiv m_1+m_2$, reduced mass $\mu\equiv m_1m_2/m$, and symmetric mass ratio ${\eta\equiv\mu/m}$, to lowest order the dimensionless orbital velocity parameter is given by
\begin{align}
    \frac{v}{c} &= \biggl(\frac{Gm}{c^3}\pi f\biggr)^{1/3},
\end{align}
\noindent where $f$ is the gravitational wave frequency defined through Kepler's third law $(\pi f)^2r^3 = Gm$. For a system to merge within the Hubble time $t_H \equiv H_0^{-1} = 14.4\text{ Gyr}$ (using the Planck parameters, \citealt{planck15}), the initial velocity of the two BHs should be (assuming $m_1=m_2$ so that $\eta=1/4$) 
\begin{eqnarray}
    \frac{v_0}{c} &=& \biggl(\frac{5}{256}\frac{1}{\eta}\frac{Gm}{c^3}\frac{1}{t}\biggr)^{-1/8} \nonumber \\
    &=& 1.74\times10^{-3}\,\eta_{0.25}^{1/8}\biggl(\frac{m}{100\,M_\odot}\biggr)^{1/8}\biggl(\frac{t}{t_H}\biggr)^{-1/8},
\end{eqnarray}
\noindent or $v_0 \simeq 522 \text{ km/s}$.

This value is much greater than the nominal value $v = 10 \, {\rm km \, s^{-1}}$ in our calculations.  However, if the orbital separation, $a$, of the binary is much smaller than the Bondi radius:
\begin{align}
    R_B &\equiv \frac{2Gm}{c_s^2}\nonumber\\
    &= 2.65\times10^{16}\,{\rm cm}\,\biggl(\frac{m}{100\,M_\odot}\biggr)c_{s,6}^{-2},
\end{align}
then the BH system will accrete at a rate similar to a system consisting of a single BH with a mass equal to the total mass of the binary; see \cite{Comerford2019} and \cite{Antoni2019}. For a binary to merge within $t_H$ in our nominal MC, it will always be true that $a\ll R_B$. 

For GW190521, there are two possibilities to allow our envisaged scenario to happen. The first possibility is that two relatively massive BHs (probably formed from first generation stars) were not initially in a binary system. After significant mass growth in an MC for a period of $<10^8$ yr, they form a binary system through dynamical interactions with other massive bodies in the MC. The binary system subsequently loses orbital energy and angular momentum through gravitational waves and merges after billions of years. The second possibility is that the two BHs were in a binary system initially within the common Bondi radius and grow together with the Bondi accretion rate. The relative speeds between the BHs and the MC are much smaller than the orbital speeds so that the nominal parameters used in our calculations can be realized. Both scenarios require relatively contrived conditions. However, in light of the evidence that GW190521-like massive BH mergers are rare, the demanding conditions imposed in our model would eliminate systems that might over-produce massive BH binaries.

\subsection{LB-1}

LB-1 is a controversial system. \cite{Liu19} initially reported this black hole candidate via radial velocity measurements of the B-type star LB-1 and estimated its mass to be $M_1 = 68_{-13}^{+11}\,M_\odot$, with a companion star mass and age of $M_2 = 8.2_{-1.2}^{+0.9}\,M_\odot$ and $t = 35_{-7}^{+13}\text{ Myr}$, respectively. \cite{AbdulMasih20} disfavored a black hole with mass $M_1>50\,M_\odot$ while favoring a lower mass companion star with $M_2 = 4.2_{-0.7}^{+0.8}\,M_\odot$. In a reply, \cite{Liu20} updated their black hole and companion star mass estimates to $M_1 \sim (23-65)\,M_\odot$ and $M_2 \sim (5-8)\,M_\odot$, respectively. In a follow-up study \cite{Liu20b} favor a mass ratio of $M_1/M_2 = 5.1\pm0.8$, thus estimating a black hole of mass $M_1 \sim (4-36)\,M_\odot$ and a companion star of mass $M_2 \sim (1-6)\,M_\odot$. The initial reported BH mass challenged the stellar evolution theories, and the scenario discussed in this paper would help to account for its existence. Though the latest upper bound on $M_1$ no longer challenges the stellar evolution theories, we suggest that the existence of $\sim 70\,M_\odot$ BHs in the Milky Way is not impossible according to the scenario discussed in this paper; but such holes should be rare given the demanding conditions discussed above.

\section{Conclusions and discussion}

We outlined a physical scenario for growing stellar-mass BHs if they are embedded in a dense MC and have slow motion with respect to the MC medium. For nominal parameters of $n = 10^5 \, {\rm cm}^{-3}$ and $v = 10 \, {\rm km \, s^{-1}}$ (insensitive to the MC temperature), we find that BHs with mass $10\,M_\odot$ and $40\,M_\odot$ can grow by $\Delta M/M \sim 10\%$ and $57\%$, respectively. We apply this scenario to GW190521, and found that two relatively massive BHs with masses $\sim 41\,M_\odot$ and $\sim 48\,M_\odot$ can grow to the observed masses from the same MC, with the condition that the relative velocities with respect to the MC for both BHs are $v \sim 10 \, {\rm km \, s^{-1}}$. Our scenario provides a novel channel for producing heavy BHs in the universe. According to this scenario, the existence of a $\sim 70\,M_\odot$ BH in the Milky Way is not impossible, even though the claimed LB-1 BH no longer needs such a large mass. Searches for such massive BHs in the Galaxy and beyond are encouraged.

\acknowledgments We thank an anonymous referee for constructive comments, Daniel Proga and Tim Waters for discussions on numerical simulations of BH accretion, and Jifeng Liu for discussion on LB-1.

\bibliography{references}
\end{document}